\documentstyle[preprint,aps,epsf,eqsecnum]{revtex}
\tightenlines
\begin{document}
\def\non{\nonumber}
\def\be{\begin{eqnarray}}
\def\en{\end{eqnarray}}
\def\la{\langle}
\def\ra{\rangle}
\def\pr{{\sl Phys. Rev.}~}
\def\prl{{\sl Phys. Rev. Lett.}~}
\def\pl{{\sl Phys. Lett.}~}
\def\np{{\sl Nucl. Phys.}~}
\def\zp{{\sl Z. Phys.}~}
\preprint{
\font\fortssbx=cmssbx10 scaled \magstep2
\hbox to \hsize{
\hfill$\raise .5cm\vtop{
\hbox{NCTU-HEP-0003}}$}
}
\draft
\vfill
\title{\Large\bf
Charge and Transition Form Factors of Light Mesons with Light-Front Quark Model
}

\vfill
\author{Chien-Wen Hwang}
\address{\rm Institute of Physics, National Chiao-Tung University,
Hsinchu, Taiwan \rm}
\vfill
\maketitle
%
\begin{abstract}
The charge and transition form factors of pion ($F_\pi$, $F_{\pi\gamma}$, and $F_{\pi\gamma^*}$) are studied with the light-front quark model. We find that our results for $F_\pi$ and $F_{\pi\gamma}$ agree very well with experimental data. Furthermore, the decay constants of $\eta$ and $\eta'$ are evaluated. We also calculate $F_{\eta\gamma}$ and $F_{\eta'\gamma}$ and compare with the experimental data.  
\end{abstract}
\pagestyle{plain}
\section{Introduction}
Form factors are very important physical quantities in understanding the internal structure of hadrons. In this paper, we study two types of form factors: charge and transition form factors for some light mesons. The former is occurred in the elastic electron meson scattering, in which one off-shell photon exchanges between the electron and one of the quarks in the meson. The latter, on the other hand, comes from the reactions where the meson is produced by one on-shell and one off-shell photons. It is well known that these form factors must be treated with the non-perturbative calculations. There are many different approaches to do that, such as lattice calculations \cite{Dan}, vector meson dominance (VMD) \cite{GL,ABBC}, perturbative QCD (pQCD) \cite{LB,LS,KO,CHM}, QCD sum rules \cite{SVZ,BH,Kho}, nonlocal quark-pion dynamics \cite{ADT,DT}, and light-front quark model (LFQM) \cite{CCC,Card,Jaus,Don,Dem,CCH}.

LFQM is the only relativistic quark model in which a consistent and fully relativistic treatment of quark spins and the center-of-mass motion can be carried out. Thus it has been applied in the past to calculate various form factors \cite{CCC,Card,Jaus,Don,Dem,CCH}. This model has many advantages. For example, the light-front wave function is manifestly boost invariant as it is expressed in terms of the momentum fraction variables (in ``+" components) in analog to the parton distributions in the infinite momentum frame. Moreover, hadron spin can also be relativistically constructed by using the so-called Melosh rotation. The kinematic subgroup of the light-front formalism has the maximum number of interaction-free generators including the boost operator which describes the center-of-mass motion of the bound state (for a review of the light-front dynamics and light-front QCD, see \cite{Zhang}). For charge and transition form factors, we concentrate on the space-like region $q^2\leq 0$ ($q$ is the momentum transfer). In this region, the so-called Z graph \cite {CCH} vanishes and only the valence-quark contributes. We take a consistent treatment with the decay constant and the charge and transition form factors in LFQM. Whatever large or small momentum transfer, it must be emphasized that these derivations are applied to all the space-like region. On the other hand, there are some experimental data whcih are concerning about the charge \cite{Amen,Bebek} and transition \cite {CLEOFpgs} form factors for some light mesons. They will offer some tests of this approach.

The paper is organized as follows. In Sec. II, the basic theoretical formalism is given and the decay constant and the charge and transition form factors are derived for pseudoscalar meson. In Sec. III,  some asymptotic behaviors and the numerical results for some light mesons are present and discussed. Finally, a summary is given in Sec. IV.

\section{Framework}
We will describe in this section the light-front approach for the
calculation of the charge and transition form factors for off-shell photons 
and light mesons. The hadronic matrix elements is 
evaluated at space-like momentum transfer, namely the region $q^2\leq 0$. 

A meson bound state consisting of a quark $q_1$ and
an antiquark $\bar q_2$ with total momentum $P$
and spin $S$ can be written as
\begin{eqnarray}
        |M(P, S, S_z)\rangle
                =\int &&\{d^3p_1\}\{d^3p_2\} ~2(2\pi)^3 \delta^3(\tilde
                P-\tilde p_1-\tilde p_2)~\nonumber\\
        &&\times \sum_{\lambda_1,\lambda_2}
                \Psi^{SS_z}(\tilde p_1,\tilde p_2,\lambda_1,\lambda_2)~
                |q_1(p_1,\lambda_1) \bar q_2(p_2,\lambda_2)\rangle,
\end{eqnarray}
where $p_1$ and $p_2$ are the on-mass-shell light-front momenta,
\begin{equation}
        \tilde p=(p^+, p_\bot)~, \quad p_\bot = (p^1, p^2)~,
                \quad p^- = {m^2+p_\bot^2\over p^+},
\end{equation}
and
\begin{eqnarray}
        &&\{d^3p\} \equiv {dp^+d^2p_\bot\over 2(2\pi)^3}, \nonumber \\
        &&|q(p_1,\lambda_1)\bar q(p_2,\lambda_2)\rangle
        = b^\dagger_{\lambda_1}(p_1)d^\dagger_{\lambda_2}(p_2)|0\rangle,\\
        &&\{b_{\lambda'}(p'),b_{\lambda}^\dagger(p)\} =
        \{d_{\lambda'}(p'),d_{\lambda}^\dagger(p)\} =
        2(2\pi)^3~\delta^3(\tilde p'-\tilde p)~\delta_{\lambda'\lambda}.
                \nonumber
\end{eqnarray}
In terms of the light-front relative momentum
variables $(x, k_\bot)$ defined by
\begin{eqnarray}
        && p^+_1=(1-x) P^+, \quad p^+_2=x P^+, \nonumber \\
        && p_{1\bot}=(1-x) P_\bot+k_\bot, \quad p_{2\bot}=x P_\bot-k_\bot,
\end{eqnarray}
the momentum-space wave-function $\Psi^{SS_z}$
can be expressed as
\begin{equation}
        \Psi^{SS_z}(\tilde p_1,\tilde p_2,\lambda_1,\lambda_2)
                = R^{SS_z}_{\lambda_1\lambda_2}(x,k_\bot)~ \phi(x, k_\bot),
\end{equation}
where $\phi(x,k_\bot)$ describes the momentum distribution of the
constituents in the bound state, and $R^{SS_z}_{\lambda_1\lambda_2}$
constructs a state of definite spin ($S,S_z$) out of light-front
helicity ($\lambda_1,\lambda_2$) eigenstates.  Explicitly,
\begin{equation}
        R^{SS_z}_{\lambda_1 \lambda_2}(x,k_\bot)
                =\sum_{s_1,s_2} \langle \lambda_1|
                {\cal R}_M^\dagger(1-x,k_\bot, m_1)|s_1\rangle
                \langle \lambda_2|{\cal R}_M^\dagger(x,-k_\bot, m_2)
                |s_2\rangle
                \langle {1\over2}s_1
                {1\over2}s_2|SS_z\rangle,
\end{equation}
where $|s_i\rangle$ are the usual Pauli spinors,
and ${\cal R}_M$ is the Melosh transformation operator:
\begin{equation}
        {\cal R}_M (x,k_\bot,m_i) =
                {m_i+x M_0+i\vec \sigma\cdot\vec k_\bot \times \vec n
                \over \sqrt{(m_i+x M_0)^2 + k_\bot^2}},
\end{equation}
with $\vec n = (0,0,1)$, a unit vector in the $z$-direction, and
\begin{equation}
        M_0^2={ m_1^2+k_\bot^2\over (1-x)}+{ m_2^2+k_\bot^2\over x}.
\label{M0}
\end{equation}
In practice it is more convenient to use the covariant form for
$R^{SS_z}_{\lambda_1\lambda_2}$ \cite{Jaus}:
\begin{equation}
        R^{SS_z}_{\lambda_1\lambda_2}(x,k_\bot)
                ={\sqrt{p_1^+p_2^+}\over \sqrt{2} ~{\widetilde M_0}}
        ~\bar u(p_1,\lambda_1)\Gamma v(p_2,\lambda_2), \label{covariant}
\end{equation}
where
\begin{eqnarray}
        &&{\widetilde M_0} \equiv \sqrt{M_0^2-(m_1-m_2)^2}, \nonumber\\
        &&\Gamma=\gamma_5 \qquad ({\rm pseudoscalar}, S=0).
\end{eqnarray}
We normalize the meson state as
\begin{equation}
        \langle M(P',S',S'_z)|M(P,S,S_z)\rangle = 2(2\pi)^3 P^+
        \delta^3(\tilde P'- \tilde P)\delta_{S'S}\delta_{S'_zS_z}~,
\label{wavenor}
\end{equation}
so that
\begin{equation}
        \int {dx\,d^2k_\bot\over 2(2\pi)^3}~|\phi(x,k_\bot)|^2 = 1. 
\label{momnor}
\end{equation}
In principle, the momentum distribution amplitude
$\phi(x,k_\bot)$ can be obtained by solving the light-front
QCD bound state equation\cite{Zhang,Cheung}.
However, before such first-principle
solutions are available, we would have to be contented with
phenomenological amplitudes.  One example that has been often
used in the literature for heavy mesons is the Gaussian-type wave function,
\begin{equation}
        \phi(x,k_\bot)_{\rm G}={\cal N} \sqrt{{dk_z\over dx}}
        ~{\rm exp}\left(-{\vec k^2\over 2\omega^2}\right),
        \label{gauss}
\end{equation}
where ${\cal N}=4(\pi/\omega^2)^{3/4}$ and $k_z$ is of the internal momentum
$\vec k=(\vec{k}_\bot, k_z)$, defined through
\begin{equation}
1-x = {e_1-k_z\over e_1 + e_2}, \qquad
x = {e_2+k_z \over e_1 + e_2},
\end{equation}
with $e_i = \sqrt{m_i^2 + \vec k^2}$. We then have 
\be
M_0=e_1 + e_2,~~~~k_z = \,{xM_0\over 2}-{m_2^2+k_\perp^2 \over 2 xM_0},
\label{kz}
\en
and
\begin{equation}
        {{dk_z\over dx}} = \,{e_1 e_2\over x(1-x)M_0}
\end{equation}
which is the Jacobian of transformation from $(x, k_\bot)$ to
$\vec k$.
This wave function has been also used in many other studies
of hadronic transitions. A variant of the Gaussian-type
wave function is
\be
\tilde {\phi}(x,k_\perp)_G=\,{\cal N}\sqrt{dk_z\over dx}\exp\left(-{M_0^2\over 2\omega^2}
\right), \label{variant}
\en
with $M_0$ being given by (\ref{M0}). This wave function is equivalent to
$\phi(x,k_\perp)_{\rm G}$ when the constituent quark masses are equal; otherwise, the results will be different. In this paper, we will assume that the $u$ and $d$ quarks in pion have the same masses. There is another wave function \cite{Jausnew}
\be
\phi(x,k_\perp)_M=\,{\cal N'}\exp\left(-{M_0^2\over 8\omega^2}\right), \label{Jausnew1}
\en
which will be also used in the numerical calculations.
\vskip 0.3cm
\subsection{Decay Constants}
The decay constant of a pseudoscalar meson $P(q_1\bar{q}_2)$ is defined by
\be
\la 0|A^\mu|P(P)\ra=\,i\sqrt{2}f_P P^\mu,
\en
where $A^\mu$ is the axial vector current. It can be evaluated by using the light-front wave function given by (2.1) and (2.5)
\be
\la 0|\bar{q}_2\gamma^+\gamma_5q_1|P\ra &=& \int \{d^3p_1\}\{d^3p_2\}
2(2\pi)^3\delta(\tilde{P}-\tilde{p_1}-\tilde{p_2})\phi_P(x,k_\perp)R^{00}_{\lambda_1\lambda_2}
(x,k_\perp)   \non \\
&& \times\,\la 0|\bar{q}_2\gamma^+\gamma_5q_1|q_1\bar{q}_2\ra.
\en
Since $\widetilde{M}_0\sqrt{x(1-x)}=\sqrt{{\cal A}^2+k^2_\perp}$,
it is straightforward to show that
\be
f_P=\,2\sqrt{3}\int {dx\,d^2k_\perp\over 2(2\pi)^3}\,\phi_P(x,
k_\perp)\,{{\cal A}\over\sqrt{{\cal A}^2+k_\perp^2}}, \label{fp}
\en
where
\be
{\cal A}=m_1x+m_2(1-x).
\en
Note that the factor $\sqrt{3}$ in (\ref{fp}) arises from the color factor
implicitly in the meson wave function. We illustrate this process in Fig.1 (a). When the decay constant is known, it can be used to constrain the parameters of the light-front wave function.
\subsection{Charge Form Factors}

The charge form factor of a pseudoscalar meson $P$ is determined by the scattering of one virtual photon and one meson. We illustrate this process in Fig.1 (b). This form factor can be defined by the matrix element
\be
\la P(P')|J^\mu|P(P)\ra = F_P(Q^2) (P+P')^\mu \label{FPdef}
\en
where $J^\mu$ is the vector current and $Q^2\equiv-q^2=-(P-P')^2$. As discussed in the above subsection, we readily obtain
\be
\la P(P')|\bar{q}\gamma^\mu q|P(P)\ra &=&\sum_{\lambda_1,\lambda'_1,\lambda_2,\lambda'_2}\int \{d^3 p_1\}\{d^3 p_2\}2(2\pi)^3\delta(p-p_1-p_2) \non \\
&&~~~~~~~~~~~~\phi^*_P(x',k'_\perp) \phi_P(x,k_\perp) R^{00\dagger}_{\lambda'_1\lambda'_2} R^{00}_{\lambda_1\lambda_2},\label{FPlf}
\en
where $k'_\perp \equiv k_\perp+x q_\perp$. After comparing (\ref{FPdef}) with (\ref{FPlf}), we can get
\be
F_P (Q^2) = \int {dx~d^2 k_\perp\over {2 (2\pi)^3}}\phi^*_P(x',k'_\perp) \phi_P(x,k_\perp) {\widetilde{M}_0\over {\widetilde{M'}_0}} \Bigg(1+{x q_\perp \cdot k_\perp \over {{\cal A}^2+k_\perp^2}}\Bigg).\label{FPcal}
\en
Thus, if we can determine the parameters in the wave function, we will have the charge form factor $F_P (Q^2)$ in terms of (\ref{FPcal}).

\subsection{Transition Form Factors}

There are two types of the transition form factors: $F_{\pi\gamma}$ and $F_{\pi\gamma*}$. The form factor $F_{P\gamma}$, in which the meson is produced by one on-shell and one off-shell photon $(\gamma\gamma^*\to P)$, is defined by the $P\gamma\gamma^*$ vertex\cite{LB}
\be
\Gamma_\mu=-ie^2~F_{P\gamma} (Q^2)\varepsilon_{\mu\nu\rho\sigma} P^\nu q^\rho \epsilon^\sigma,
\en
where $q$ is the momentum of the off-shell photon, $q^2=-q^2_\perp=-Q^2$, and $\epsilon$ is the polarization vector of the on-shell photon. We illustrate this process in Fig.1 (c) and write down the amplitude in the light-front framework \cite{LB}
\be
\Gamma_\mu &=& \sum_{\lambda_1,\lambda_2,\lambda} e_q e_{\bar q'}e^2\int \{d^3 p_1\}\{d^3 p_2\}2(2\pi)^3\delta(\tilde{P}-\tilde{p_1}-\tilde{p_2}) \phi_P(x,k_\perp) \non \\
&\times& \Bigg[\Bigg({{q^2_\perp\over {p^+}} - {m_1^2+(k_\perp+q_\perp)^2\over{p_1^+}} - {m_2^2+k^2_\perp\over{p_2^+}}}\Bigg)^{-1} \bar {v}(p_2,\lambda_2)\not{\!\epsilon} u(p'_1,\lambda)\bar {u}(p'_1,\lambda)\gamma_\mu u(p_1,\lambda_1) \non \\
&&+(1 \leftrightarrow 2)\Bigg]R^{00}_{\lambda_1\lambda_2},
\en
where $e_q$ and $e_{\bar q'}$ are the electric charge of $q$ and ${\bar q'}$ quarks, respectively. It is straightforward to show that
\be
F_{P\gamma}(Q^2) &=& -4{\sqrt{3}\over{\sqrt{2}}}e_q e_{\bar q'}\int {dx\,d^2k_\perp\over 2(2\pi)^3}\,\phi_P(x,
k_\perp)\,{{\cal A}\over\sqrt{{\cal A}^2+k_\perp^2}} \non \\
& \times &\Bigg[{1\over{(1-x) \left(q^2_\perp-{m^2_1+(k_\perp+q_\perp)^2\over{1-x}}-{m^2_2+k^2_\perp\over{x}}\right)}}+{1\over{x \left(q^2_\perp-{m^2_1+k_\perp^2\over{1-x}}-{m^2_2+(k_\perp-q_\perp)^2\over{x}}\right)}}\Bigg]\label{FMG} 
\en
The form factor $F_{P\gamma^*}$ arising from  the $P\gamma^*\gamma^*$ vertex, where $\gamma*\gamma*$ represents two off-shell photons is defined by \cite{KWZ}
\be
\Gamma_{\mu\nu} = -ie^2F_{P\gamma^*}(Q^2,Q'^2) \varepsilon_{\mu\nu\rho\sigma} {\cal Q}^\rho {\cal P}^\sigma, \label{229}
\en
where ${\cal Q}\equiv {1\over{2}}(q'-q)$, ${\cal P}\equiv q'+q$, and $Q'^2=-q'^2=q'^2_\perp$. This process is illustrated in Fig.1 (d). We expect that this amplitude is similar to the one off-shell photon case and we have
\be
\Gamma_{\mu\nu} &=& \sum_{\lambda_1,\lambda_2,\lambda} e_q e_{\bar q'}e^2\int \{d^3 p_1\}\{d^3 p_2\}2(2\pi)^3\delta(p-p_1-p_2) \phi_P(x,k_\perp) \non \\
&\times& \Bigg[\Bigg({{q^2_\perp-q'^2_\perp\over {p^+}} - {m_1^2+(k_\perp+q_\perp)^2\over{p^+_1}} - {m_2^2+k^2_\perp\over{p^+_2}}}\Bigg)^{-1} \bar {v}(p_2,\lambda_2)\gamma_\nu u(p'_1,\lambda)\bar {u}(p'_1,\lambda)\gamma_\mu u(p_1,\lambda_1) \non \\
&&+(1 \leftrightarrow 2)\Bigg]R^{00}_{\lambda_1\lambda_2}. \label{230}
\en 
From (\ref{229}) and (\ref{230}), we arrive at
\be
F_{P\gamma^*}(Q^2,Q'^2) &=& -4{\sqrt{3}\over{\sqrt{2}}}e_q e_{\bar q'}\int {dx\,d^2k_\perp\over 2(2\pi)^3}\,\phi_P(x,
k_\perp)\,{{\cal A}\over\sqrt{{\cal A}^2+k_\perp^2}} \non \\
& \times &\Bigg[{1\over{(1-x) \left(q^2_\perp-q'^2_\perp-{m^2_1+(k_\perp+q_\perp)^2\over{1-x}}-{m^2_2+k^2_\perp\over{x}}\right)}} \non \\
&&+{1\over{x \left(q^2_\perp-q'^2_\perp-{m^2_1+k_\perp^2\over{1-x}}-{m^2_2+(k_\perp-q_\perp)^2\over{x}}\right)}}\Bigg]. \label{FPscal}
\en
\section{Numerical Results and Discussions}
We now compare our results of the form factors with the experimental data. Before doing that, we must determine the parameters $m_1$, $m_2$, and $\omega$ in the wave function $\phi_P(x,k_\perp)$. Of course, we assume that this wave function is process-independent.

In the $\pi-\gamma$ case, the constituent masses of the $u$ and $d$ quarks are the same, i.e., $m_1=m_2\equiv m_q$. We can use the experimental value of decay constant $f_\pi = 92.4$ MeV\cite{PDG98} to determine the parameters $m_q$ and $\omega_\pi$ by (\ref{fp}). However, there are two parameters with only one experimental value. Therefore, principally, we can get infinite combinations which all satisfy the decay constant value. If we can find another constraint, these parameters wull be deterimined uniquely. From the transition form factor $F_{\pi\gamma}$, we have another constraint. In (\ref{FMG}), if we consider the limit $Q^2 \to \infty$, there is a simple form
\be
F_{\pi\gamma}(Q^2 \to \infty) = 4{\sqrt{3}\over{\sqrt{2}}}{(e_u^2-e_d^2)\over{\sqrt{2}}}\int {dx\,d^2k_\perp\over 2(2\pi)^3}\,\phi_\pi(x,
k_\perp)\,{{\cal A}\over\sqrt{{\cal A}^2+k_\perp^2}}\left({1\over{x (1-x) Q^2}}\right). \label{Fpinfty}
\en
From \cite{LB}, we have
\be
Q^2 F_{\pi\gamma}(Q^2)\Bigg\vert_{Q^2\to\infty}=6(e_u^2-e_d^2)f_\pi. \label{LBFp}
\en
Comparing (\ref{Fpinfty}) with (\ref{LBFp}), we obtain
\be
f_\pi={\sqrt{3}\over{3}} \int {dx\,d^2k_\perp\over 2(2\pi)^3}\,{\phi_\pi(x,
k_\perp)\over{x (1-x)}}{{\cal A}\over\sqrt{{\cal A}^2+k_\perp^2}}.\label{fp2}
\en
From (\ref{fp}) and (\ref{fp2}), we can uniquely determine all the parameters in the wave function by using only one experimental value $f_\pi$. Here we show the parameters of two wave functions $\phi_G$ and $\phi_M$ fitted to the decay constants given by (\ref{fp}) and (\ref{fp2})($\phi_G=\tilde{\phi}_G$ in pion case) as
\be
\phi_G&:&~~m_g=0.243~\text{GeV},~~\omega_\pi=0.328~\text{GeV};\\
\phi_M&:&~~m_g=0.198~\text{GeV},~~\omega_\pi=0.513~\text{GeV}.
\en
We use the Gaussian-type wave function to calculate the form factors because the value $m_q$ of the wave function $\phi_M$ seems to be too small. Moeover, we can evaluate $F_\pi(Q^2)$ and $F_{\pi\gamma}(Q^2)$ in all momentum transfer region by using (\ref{FPcal}), (\ref{FMG}). Because both parameters which we needed have been fixed, we have no degree of freedom to adjust this wave function. Thus, whether these preditions are consistent with the experiments or not are very strict tests for the Gaussian-type model. From Fig.2 and Fig.3, we find that these predictions are in good agreement with the experimental data \cite{Amen,Bebek,CLEOFpgs}. 

For $F_{\pi\gamma^*}(Q^2,Q'^2)$, since there are no experimental data yet, we must proceed carefully. If we take both limits of $Q^2,Q'^2 \to \infty$, (\ref{FPscal}) becomes
\be
F_{\pi\gamma^*}(Q^2,Q'^2)\Bigg\vert_{Q^2,Q'^2 \to\infty} &=& {2\over{\sqrt{3}}}\int {dx\,d^2k_\perp\over 2(2\pi)^3}\,\phi_\pi(x,
k_\perp)\,{{\cal A}\over\sqrt{{\cal A}^2+k_\perp^2}}\non \\
&&\times\left({1\over{x Q^2+ (1-x) Q'^2}}+{1\over{x Q'^2+ (1-x) Q^2}}\right). \label{Fpsinfty2} 
\en
Noting that, the wave function $\phi_\pi (x,k_\perp)$ is symmetric in $x$ and $1-x$, we get the asymptoyic behavior of the transition form factor as
\be
F_{\pi\gamma^*}(Q^2,Q'^2)\Bigg\vert_{Q^2,Q'^2 \to\infty} = {4\over{\sqrt{3}}}\int {dx\,d^2k_\perp\over 2(2\pi)^3}\,\phi_\pi(x,
k_\perp)\,{{\cal A}\over\sqrt{{\cal A}^2+k_\perp^2}}\left({1\over{x Q^2+ (1-x) Q'^2}}\right), \label{Fpsinfty} 
\en
which is consistent with the assumption in \cite{KO}. Thus we have the confidence to make the prediction about the values of $F_{\pi\gamma^*}(Q^2,Q'^2)$ in terms of (\ref{FPscal}).

We also consider the $\eta -\eta'$ system. Due to the mixing in this system, $\eta$ and $\eta'$ both have $\eta_8$ and $\eta_0$ components. These two states
\be
|\eta_8\ra &=& {1\over{\sqrt{6}}}|\bar {u}u+\bar {d}d-2\bar {s}s\ra \non 
\en
and
\be
|\eta_0\ra &=& {1\over{\sqrt{3}}}|\bar {u}u+\bar {d}d+\bar {s}s\ra  \non 
\en
are the $SU(3)$ octet and singlet, respectively. The decay constants of octet and singlet, $f^a_P$ are defined as 
\be
\la 0|J^a_{\mu 5}|P(p)\ra = {\sqrt {2}}if^a_P~p_\mu, ~~~(a=8,1;~~P=\eta,\eta'),
\en
where $J^a_{\mu 5}$ denotes axial-vector current. Recent investigations \cite{Leu,FK,FKS} have shown that the decay constants of the $\eta-\eta'$ system are not adequately described by only one mixing angle. A two-mixing-angle parametrization is given by \cite{FKS,CCTY}
\be
&&f^8_\eta=f_8 \cos \theta_8,~~~f^0_\eta=-f_0 \sin \theta_0, \non \\
&&f^8_{\eta'}=f_8 \sin \theta_8,~~~f^0_{\eta'}=f_0 \cos \theta_0,
\en
where $\theta_8 \neq \theta_0$. From the phenomenological analysis \cite{Feld}, we get the values
\be
&&f_8 \simeq 1.26 f_\pi, ~~~\theta_8 \simeq -21.2^\circ, \non \\
&&f_0 \simeq 1.17 f_\pi, ~~~\theta_0 \simeq -9.2^\circ.  \label{fff}
\en
Using the values of $f_8$ and $f_0$, we can deterimine the parameters by (\ref{fp}) and (\ref{fp2})
\be
m_8=0.306~\text{GeV}~~\omega_{\eta_8}=0.414~~\text{GeV}, \non \\
m_0=0.285~\text{GeV}~~\omega_{\eta_0}=0.384~~\text{GeV}, \label{310}
\en
where $m_8$ and $m_0$ are the parameters of quark masses in $\eta_8$ and $\eta_0$, respectively. With these parameters in (\ref{310}) and the values of mixing angles in (\ref{fff}), we could calculate $F_{\eta\gamma}$ and $F_{\eta'\gamma}$ and the results are ploted in Fig.4.


\section{Summary}

The charge and transition form factors of $\pi$ have been studied in the present paper. In the light-front relativistic quark model, these form factors have been evaluated in a frame where $q^+=0$ and $q^2\leq 0$ and there is no need to calculate the contribution from the so-called $Z$ graph \cite{CCH}. 

We have only used one experimental value, the pion decay constant, to fix the two parameters in the pion wave function. This point is in contrast to pQCD which treats the decay constant as one part of the wave function. Thus, we would emphasize that the wave function contains no more degree of freedoms to adjust. When the parameters are fixed, we evaluate the charge as well as one and two virtual photon transition form factors in $-8~{\text GeV}^2\le q^2 <0$. Our calculations are based on the important assumption that: the wave functions are independent of processes. We compare the results of calculation with the experimental data and find that this assumption is valid for $-8~{\text GeV}^2\le q^2 <0$ region. Basing on these consistency of $F_\pi$ and $F_{\pi\gamma}$, we have the confidence to make the prediction of the $F_{\pi\gamma^*}$. The decay constants $f_{\eta_8}$, $f_{\eta_0}$, and the mixing angles $\theta_8$, $\theta_0$ have been obtained by using the phenomenological analysis. With the same approach, we have gotten $F_{\eta\gamma}$ and $F_{\eta'\gamma}$ which agree well with the experimental data.

\vskip 2.0cm
\centerline {\bf ACKNOWLEDGMENTS}

   We are grateful to T. Feldmann and H.Y. Cheng for the helpful discussions on the $\eta-\eta'$ mixing system. This work was supported in part by the National Science Council of ROC under Contract Nos. NSC89-2112-M-009-035. 

%
%
\newcommand{\bi}{\bibitem}
%

\newpage
\parindent=0 cm
\centerline{\bf FIGURE CAPTIONS}
\vskip 0.5 true cm

{\bf Fig. 1 } The diagram of (a) one pseudoscalar meson decay to vacuum, (b) the scattering of one virtual photon and one meson, (c) a meson is produced by one on-shell and one off-shell photons, and (d) a meson is produced by two off-shell photons.
\vskip 0.25 true cm

{\bf Fig. 2 } The charge form factor of pion in small and large momentum transfer. Data are taken from \cite{Amen} and \cite{Bebek}, respectively.
\vskip 0.25 true cm

{\bf Fig. 3 } The one off-shell photon transition form factor of pion. The solid line represents the results obtained with this approach. The dotted line represents the limiting behavior $2 f_\pi$ ($0.185$ GeV). Data are taken from \cite{CLEOFpgs}.
\vskip 0.25 true cm

{\bf Fig. 4 } The one off-shell photon transition form factor of $\eta$ and $\eta'$. The dotted line represents the limiting behavior $0.182$ GeV and $0.300$ GeV, respectively. Data are both taken from \cite{CLEOFpgs}. 
\vskip 0.25 true cm

\newpage

\begin{figure}[h]
\hskip 4cm
\hbox{\epsfxsize=16cm
      \epsfysize=20cm
      \epsffile{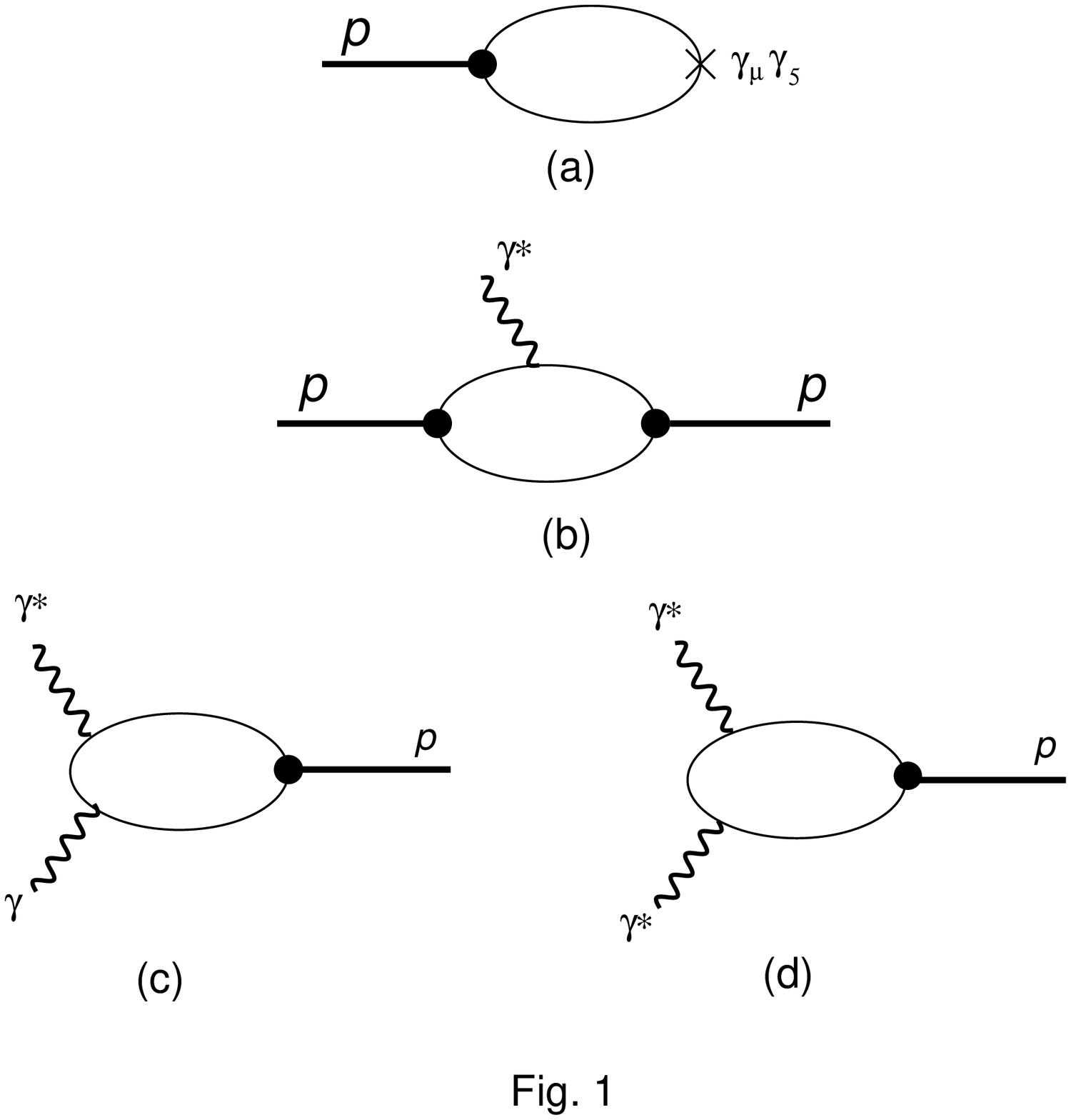}}
\end{figure}
\newpage

\begin{figure}[h]
\hskip 4cm
\hbox{\epsfxsize=16cm
      \epsfysize=20cm
      \epsffile{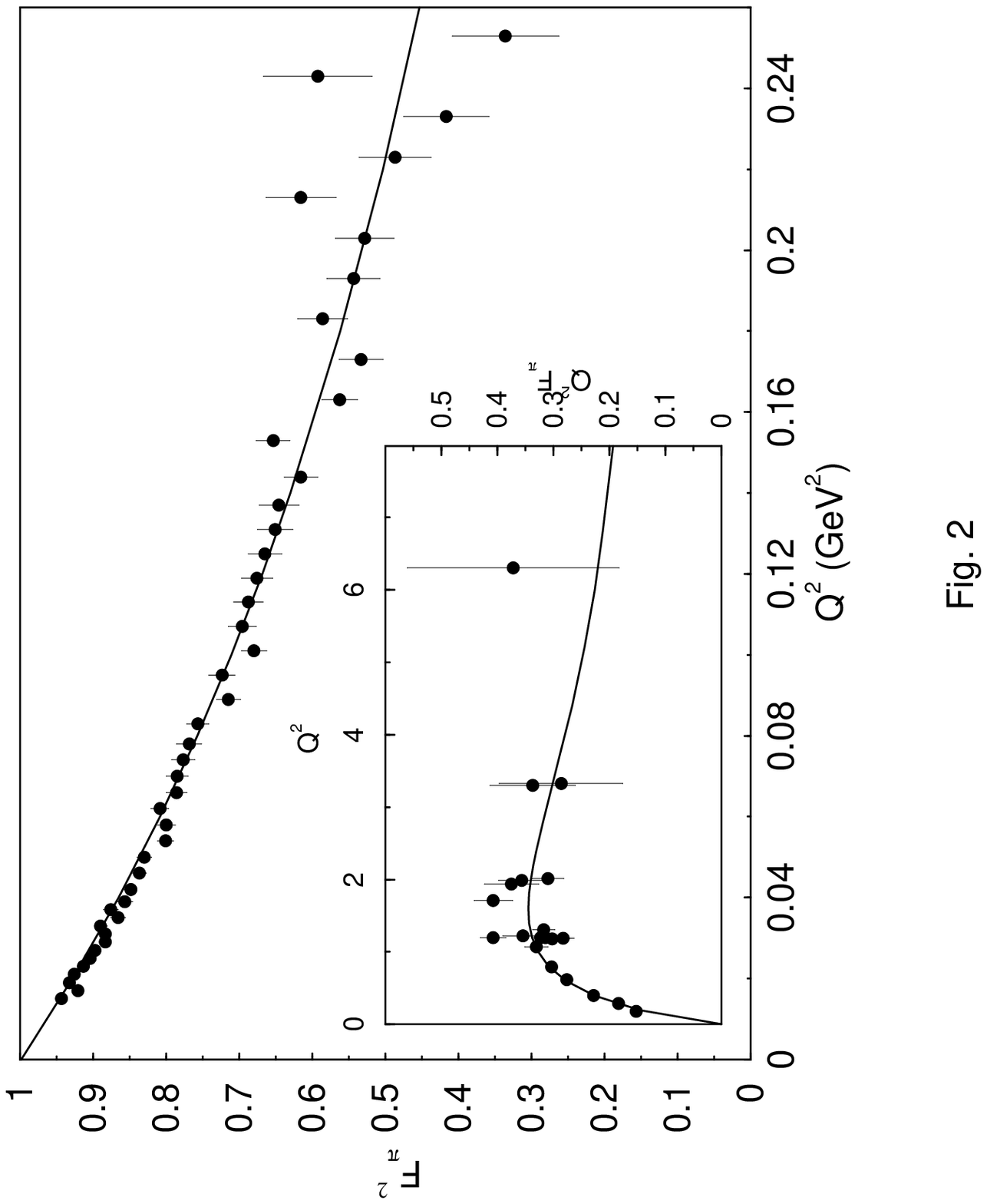}}
\end{figure}

\begin{figure}[h]
\hskip 4cm
\hbox{\epsfxsize=16cm
      \epsfysize=20cm
      \epsffile{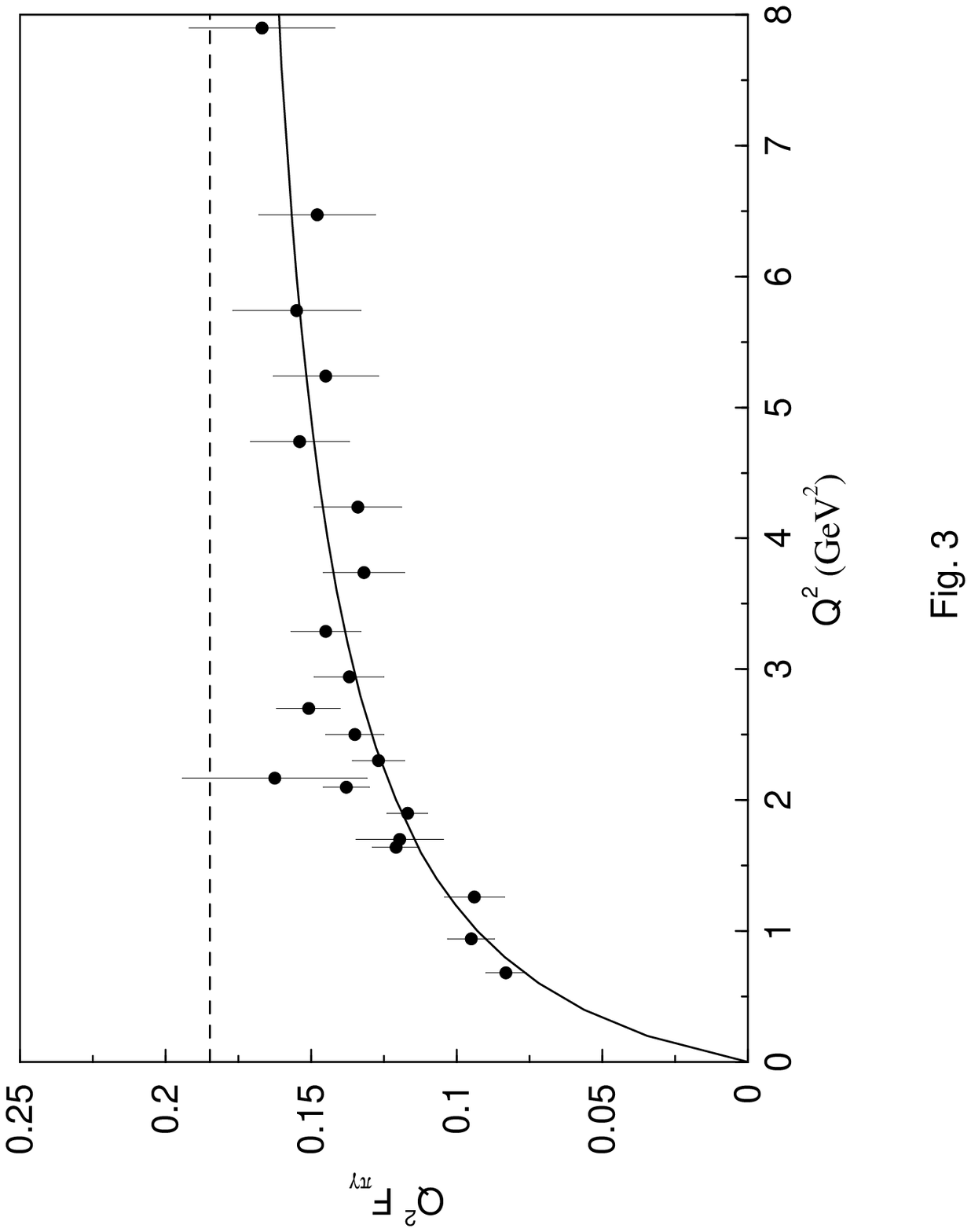}}
\end{figure}
\newpage

\begin{figure}[h]
\hskip 4cm
\hbox{\epsfxsize=16cm
      \epsfysize=20cm
      \epsffile{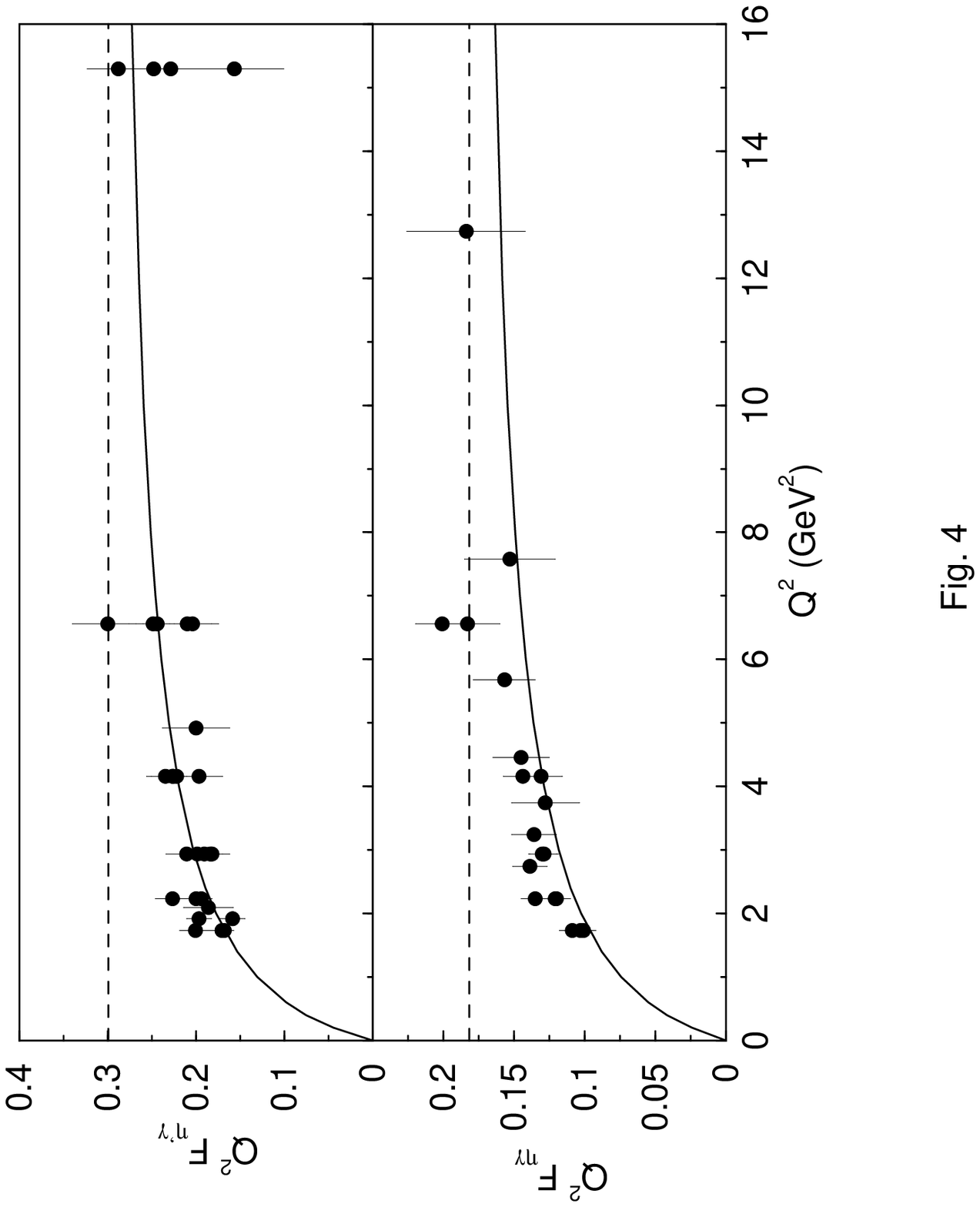}}
\end{figure}
\newpage

\end{document}